\documentclass[aps,pra,superscriptaddress,twocolumn, 10pt]{revtex4-1}
\usepackage{graphicx}
\usepackage{bm}
\usepackage[usenames]{color}
\bibstyle{apsrev.bib}

\usepackage{epsfig}
\usepackage{amsmath}
\DeclareMathOperator{\Tr}{Tr}
\usepackage{amssymb}
\usepackage{array}
 
\newcommand{\be}{\begin{equation}}
\newcommand{\ee}{\end{equation}}
\newcommand{\beqn}{\begin{eqnarray}}
\newcommand{\eeqn}{\end{eqnarray}}

\usepackage{dsfont}

\begin{document}

\title{Entanglement scaling in fermion chains with a localization-delocalization transition and inhomogeneous modulations}

\author{Gerg\H o Ro\'osz}
\email{roosz.gergo@wigner.hu}
\affiliation{Institute of Theoretical Physics, Technische Universit\"at Dresden, 01062 Dresden, Germany}
\affiliation{Wigner Research Centre for Physics, Institute for Solid State Physics and Optics, H-1525 Budapest, P.O. Box 49, Hungary}

\author{Zolt\'an Zimbor\'as}
\email{zimboras.zoltan@wigner.hu}
\affiliation{Wigner Research Centre for Physics, Theoretical Physics Department, H-1525 Budapest, P.O. Box 49, Hungary}
\affiliation{MTA-BME Lend\" ulet Quantum Information Theory Research Group}
\affiliation{Mathematical Institute, Budapest University of Technology and Economics, Hungary}

\author{R\'obert Juh\'asz}
\email{juhasz.robert@wigner.hu}
\affiliation{Wigner Research Centre for Physics, Institute for Solid State Physics and Optics, H-1525 Budapest, P.O. Box 49, Hungary}


\date{\today}

\begin{abstract}
We study the scaling of logarithmic negativity between adjacent subsystems 
in critical fermion chains with various inhomogeneous modulations through
numerically calculating its recently established lower and upper bounds. 
For random couplings, as well as for a relevant aperiodic modulation of the couplings, which induces an aperiodic singlet state, the bounds are found to increase logarithmically with the subsystem size, and both prefactors agree with the predicted values characterizing the corresponding asymptotic singlet state. For the marginal Fibonacci modulation, the prefactors in front of the logarithm are different for the lower and the upper bound, and vary smoothly with the strength of the modulation.    
In the delocalized phase of the quasi-periodic Harper model, the scaling of the bounds of the logarithmic negativity as well as that of the entanglement entropy are compatible with the logarithmic scaling of the homogeneous chain. 
At the localization transition, the scaling of the above entanglement characteristics holds to be logarithmic, but the prefactors are significantly reduced compared to those of the translationally invariant case, roughly by the same factor.  
\end{abstract}

\maketitle


\section{Introduction}

In the last decade, there has been an increasing interest in the  the entanglement properties of quantum systems \cite{eisert2008,entanglement_review,laflorencie}. The studies on this subject allowed for a deeper understanding of many-body models, in particular with respect to criticality \cite{vidal,Calabrese_Cardy04, hsu2009}, simulability \cite{VC2006, shi2006, SWVC2008}, and thermalization \cite{EFG2015, AC2017, serbyn2019}. In many of the early works, the case when the entire quantum system is in a pure state was considered, and one investigated the entanglement between a subsystem and the rest of the system.
In this case, the so-called entanglement entropy is a proper measure of the quantum entanglement \cite{ryszard2009}.  Consider a system $S$ in a pure state $| \Psi \rangle$, and divide the system into two complementary parts $A$ and $B$ (i.e., in a way that $A \cup B = S$); calculating the reduced density matrices $\rho_{A} = \Tr_{B} | \Psi \rangle \langle \Psi |$,  $\rho_{B} = \Tr_{A} | \Psi \rangle \langle \Psi |$,  the entanglement entropy is  defined as the von Neumann entropy of either of them:
\begin{equation}
S = -\Tr_A \rho_A \ln \rho_A = -\Tr_B \rho_B \ln \rho_B  \;.
\end{equation}

However, if one wishes to characterize the entanglement between two subsystems whose union does not cover the whole system $S$, ($A \cup B \neq S$), the $A \cup B$ subsystem is no longer in a pure state, and one has to use another measure to quantify the entanglement between $A$ and $B$.
One possible candidate is the logarithmic entanglement negativity, which is proven to be an entanglement monotone \cite{eisert2001,vidal2002} and is defined as
\begin{equation}
\mathcal{E}_N = \ln || \rho^{T_A}_{A \cup B}||_1 \;, 
\label{log-ent-neg}
\end{equation}  
where $\rho_{A \cup B}$ is the reduced density matrix of the subsystem  $A \cup B$, $T_A$ denotes the partial transpose on subsystem $A$, and $||\cdot||_1$ is the trace norm. 
The partial transpose is defined as $\langle \varphi_A \varphi_B| \rho^{T_A}_{A \cup B} | \varphi'_A \varphi'_B  \rangle  = \langle \varphi'_A \varphi_B| \rho_{A \cup B} | \varphi_A \varphi'_B  \rangle$, with $\langle \varphi_A |$ and $\langle \varphi_B |$ being bases  for $A$ and $B$, respectively. It corresponds to the local time reversal, and if a state is separable (non-entangled) partial transposed density matrix is still a valid density matrix, while if it is entangled, the partial transpose of the density matrix may have negative eigenvalues \cite{peres1996}.
For one-dimensional, homogeneous critical models, there exist conformal field theory (CFT) results for the entanglement entropy and the entanglement negativity \cite{calabrese2012}.
The entanglement entropy depends on the length $\ell$ of the subsystem as  
\begin{equation}
S(\ell)= \frac{c}{3} \ln \ell + \mathrm{const} \;,
\end{equation}
where $c$ is the so-called central charge, while the entanglement negativity of 
adjacent subsystems of length $\ell$ scales as 
\begin{equation}
\mathcal{E}(\ell)=\frac{c}{4} \ln \ell  + \mathrm{const} \;.
\label{eq:lnl}
\end{equation}
In the recent years, a series of numerical \cite{bayat2010,calabrese2013,ruggiero2016,alba2013,chung2014,sherman2016,gray2018} and analytical \cite{calabrese2012,tonni2013,eisler2014,calabrese2015,coser2014,hoogeven2015,wen2015,blondeau2016, santos2011,chang2016, grover2019, Lu2019, shapourian2019} works have been devoted to the entanglement negativity.  
For non-interacting  bosonic systems, the entanglement negativity is calculable efficiently (in the number of nodes) both for the ground state and for thermal states \cite{vidal2002,audenaert2002,eisert2003,adesso2007}.
The reason  is, that the partial transpose maps the bosonic Gaussian states to bosonic Gaussian operators \cite{peres1996,simon2000}.
However, for non-interacting fermionic systems, no similar calculation method is known. The ground and thermal states are fermionic Gaussian states, but the partial transpose is not a fermionic Gaussian operator \cite{eisler2015}.
The lack of a simple formula for the entanglement negativity triggered a series of works on upper and lower bounds \cite{shapourian2017, shapourian2019, shapurian2019RevA, eisler2018, negativity-bounds}. 

In this work, we will use the upper and lower bounds given in \cite{eisler2018} to investigate numerically the scaling of the entanglement negativity in various fermionic chains with random or aperiodic inhomogeneities. In the case of sublattice-symmetry, we will present a simplification in the calculation of the lower bound. The scaling of entanglement entropy will also be studied numerically by the correlation matrix method \cite{peschel-2003}.

The rest of the paper is organized as follows. In Sec.  \ref{models}, the models are defined. In \ref{sec:neg}, we recapitulate the steps of calculating the negativity upper and lower bounds introduced in Ref.~\cite{negativity-bounds}, and a simplification of the form of the lower bound due to sublattice symmetry is presented in Sec. \ref{sec:simple}. In Sec. \ref{results}, we present our numerical results. These are then discussed in Sec. \ref{concl}.  .


\section{Models}
\label{models}
 
In this work, we will consider different variants of the fermionic hopping model having the Hamiltonian of general form
 \begin{equation}
 H=-\frac{1}{2}\sum_{l=1}^L t_l (c_{l+1}^{\dagger} c_l + c_l^{\dagger} c_{l+1}) + h_l c^{\dagger}_l c_l ,
\label{hamilton}
\end{equation}    
where $t_l$ and $h_l$ are site-dependent hopping amplitudes and on-site energies, respectively, while $c_l^\dagger$ and $c_l$ are fermionic creation and annihilation operators obeying the anticommutation rules $\{ c_l, c_j\}=\{ c_l^\dagger,c_j^\dagger\}=0$ and $\{ c_l, c_j^\dagger \} = \delta_{l,j}$, for $l,j=1,2,\dots,L$. Periodic boundary condition is considered, so that site $L+1$ is identified with site $1$. The chemical potential is zero, so the states with $E_n<0$ are occupied.
We mention that this class of models can be mapped to XX spin-$1/2$ chains by the well-known Jordan-Wigner transformation.

\subsection{Off-diagonal inhomogeneity}

For this class of models, the on-site energies are all zero $h_j=0$, $j=1,\dots,L$, while the hopping amplitudes are position-dependent, either random or follow an aperiodic modulation. 
In the first case, which we refer to as random model, we assume that the amplitudes $t_j$ are independent, identically distributed quenched random variables, drawn from a uniform distribution in the interval $[0,1]$. 

In the latter case, we will use two different aperiodic modulations, both defined by an inflation rule.  
One of them  is the Fibonacci modulation, where hopping amplitudes are modulated according to the Fibonacci sequence.
It is defined using a two-letter alphabet ($a$ and $b$) by the substitution rule:
\begin{eqnarray}
 a &\to& ab \nonumber \\
 b &\to& a  \nonumber \;.
\end{eqnarray}
The first few realizations of the Fibonacci sequence are $a$, $ab$, $aba$, $abaab$.

Another two-letter sequence, which we refer to as relevant aperiodic modulation (RAM), is obtained by the following inflation rule \cite{jz}: 
\begin{eqnarray}
 a &\to& ababa \nonumber \\
 b &\to& a  \nonumber \;.
 \label{ram}
\end{eqnarray}

Generating a sufficiently long aperiodic sequence by the repeated application of the inflation rule, a modulation pattern can be associated with it, in which letter $a$ ($b$) corresponds to amplitude $t_a$ ($t_b$).  
The strength of the modulation can be characterized by the ratio of two types of amplitudes, $r=t_a/t_b$.

 The non-zero energy eigenstates of the randomly disordered model are exponentially localized, and the localization length diverges	for zero energy as it was shown with rigorous tools in connection with the off-diagonal Anderson model in \cite{deylon1986,igloi2017}.
According to the strong-disorder renormalization group (SDRG) method \cite{mdh,im}, the ground state of the random model is a random-singlet state \cite{fisherxx},
which is a product of one-particle states $\frac{1}{\sqrt{2}}(|10\rangle -|01\rangle)$ on pairs of sites which can be arbitrarily far away from each other. The method is approximative but is asymptotically exact, giving the low-energy (large-scale) properties of the system correctly. 
Analogous to this, the ground state of aperiodic models is an aperiodic-singlet state \cite{hida}, for any $r<1$ in the case of the RAM \cite{jz}, but, for the Fibonacci modulation, which is a so-called marginal perturbation, only in the limit $r\to 0$.    

In such a singlet state, the entanglement entropy in units of $\ln 2$ is simply given by the number of pairs with precisely one constituent in subsystem $A$. The average entanglement entropy of a subsystem of size $\ell$, which is part of an infinite system increases asymptotically (apart from log-periodic oscillations for aperiodic models) as 
\begin{equation}
S(\ell)= \frac{c_{\rm eff}}{3}\ln\ell+{\rm const},
\end{equation} 
where the effective central charge depends on the type of modulation. 
For the random model\cite{refael}, $c^{\rm ran}_{\rm eff}=\ln 2$ , for the RAM, 
$c^{\rm RAM}_{\rm eff}=\frac{6 \ln 2}{ \ln \lambda} \frac{(\lambda-3)^2}{2+(\lambda-3)^2}$, where \cite{jz} $\lambda=\frac{1}{2}(3+\sqrt{17})$, while for the Fibonacci modulation it varies continuously with $r$, and its limiting value is 
$\lim_{r\to 0} c^{\rm FM}_{\rm eff}(r)=\frac{2}{(\tau^2+1) \log_2 \tau}$, where   
$\tau=\frac{1+\sqrt{5}}{2}$ is the golden mean \cite{jz,ijz}.

The logarithmic negativity in a singlet state is given by the number of singlets  connecting $A$ and $B$, in units of $\ln 2$. 
As it was shown in Ref. \cite{ruggiero2016} for the random singlet state, the average logarithmic negativity of adjacent intervals of size $\ell$ scales as 
\be
\mathcal{E}(\ell)= \kappa\ln \ell + {\rm const}\;,
\label{neg} 
\ee
with the prefactor $\kappa=\frac{\ln 2}{6}$. Recently, a more detailed study has appeared about the negativity spectrum of this model \cite{turkeshi2019}. 
The prefactor $\kappa$ is the half of the prefactor of entanglement entropy, which can be intuitively understood since subsystem $A$ borders to $B$ only on one side, while, in the case of entanglement entropy, $A$ borders to the rest of the system on both sides.     
According to this, the result in Eq. (\ref{neg}) is expected to hold also for aperiodic singlet states with a prefactor $\kappa=\frac{c_{\rm eff}}{6}$. 

The recapitulated SDRG method and its modifications has a wide range of applications from the description of entanglement of star-graph like systems  \cite{juhasz_ober_2018} and disordered surfaces \cite{juhasz2017, juhasz2017_jstatmech} including the dynamics of disordered systems \cite{pekker2014, vosk2013, vosk2014, bardason2012, roosz2017} and description of their highly excited states \cite{pouravarni2013, pouravarni2015, huang2014, vasseur2015, storms2014, hsin2015}  or even a construction of quasi-periodic tensor network model enable us to investigate AdS/CFT correspondence \cite{jahn2019}.

\subsection{The Harper model}

Another model, which we will consider is the Harper model. Here, the couplings in the general Hamiltonian in Eq. (\ref{hamilton}) are constant, $t_i=1$, while the local potential $h_j$ is a quasiperiodic function of $j$,
namely,
\be 
h_j=h \cos (2 \pi j/\tau), 
\ee
where $\tau = \frac{1+\sqrt{5}}{2}$ is the golden mean. The irrationality of $\tau$ makes the model quasiperiodic.
This model, when formulated in terms of spin variables is also known as Aubry-Andr\'e model \cite{aubry-andre}.
It is well-known that this model shows a delocalization transition, which is exactly at $h=1$ due to its self-dual property \cite{wilkinson}. 
In the region $h<1$, all one-particle eigenstates are extended, whereas, 
for $h<1$, they are all localised on a length \cite{aubry-andre}
\begin{equation}
 l_{\rm loc}=\frac{1}{\ln h} \;.
\end{equation}
At the critical point, $h=1$, the one-particle states show an interesting multi-fractal behaviour \cite{evangeleou2000,siebesma1987}: they are essentially localised on $\sim L^{D_2(n)}$ sites, where the dimension $D_2(n)$ of the effective support of the state varies from state to state. Its maximal value is found numerically\cite{evangeleou2000} to be $D_2 \approx 0.82 $ . 

To our knowledge the logarithmic negativity in the ground state of the Harper model was not studied so far in the literature, the entanglement entropy has been studied in \cite{harper_ent}, there the authors focused on the effect of the chemical potential. Here we investigate the size dependence of the entanglement entropy.

\section{Negativity of free fermions: upper and lower bounds}
\label{sec:neg}

In this section, we briefly summarize the calculation of upper and lower bounds introduced in Ref. \cite{eisler2018} and used in the present paper.
Instead of the general definitions, here, it is sufficient to restrict ourselves to the formulation valid in the special case of particle number conserving states. The latter means that the density matrix commutes with the particle number operator ($N=\sum_{l=1}^L c^{\dagger}_l c_l $),  $[\rho, N]=0$. The ground states of the quadratic fermion Hamiltonian in Eq. (\ref{hamilton}) are such states.

\subsection{Upper bound}

First, we consider the upper bound $\mathcal{E}_u$. It is formulated in terms of the covariance matrix
\begin{equation}
\gamma_{2i-1, 2j}=-\gamma_{2j, 2i-1} =2C_{ij} - \delta_{ij} \, ,
\end{equation} 
where $C_{ij}=\langle c^\dag_i c^{\phantom{\dagger}}_j\rangle$ is the correlation matrix, with all the other entries of $\gamma$ being zero.
For every Hamiltonian in the form (\ref{hamilton}), the covariance matrix can be obtained following a standard canonical transformation \cite{lieb61}. For translational invariant systems, one can easily obtain a closed form of the covariance matrix while, for inhomogeneous systems, it is computable in polynomial time in the number of fermionic modes $L$. The covariance matrix characterizes all correlations in the system, so it also determines the entanglement negativity of any subsystems. However, no simple formula is known for the entanglement negativity in terms of the covariance matrix, making the bounds defined in \cite{eisler2018} really valuable. 
The upper bound we use is given by
\begin{equation}
\mathcal{E}_u = \frac{1}{2} \left[ S_{1/2}(\rho_{\times})-S_2(\rho_{A \cup B}) \right] \;,
\end{equation}
where $S_{\alpha}(\rho)$ denotes the R\'enyi entropy of state $\rho$:
\begin{equation}
S_{\alpha}(\rho) = \frac{1}{1-\alpha} \ln \Tr \rho^\alpha \;.  
\end{equation}
The state $\rho_{\times}$ is defined by its covariance matrix as
\begin{equation}
\gamma_{\times}= \mathds{1}-(\mathds{1}-\gamma_{-})(\mathds{1}+\gamma_{+} \gamma_{-})^{-1}(\mathds{1}-\gamma_{+})\;,
\label{eq:gamma-x}
\end{equation}
where $\gamma_{\pm} = T^{\pm}_{B} \, \gamma \, T^{\pm}_{B}$, and 
$T^{\pm}_{B} =\bigoplus_{j\in A}\mathds{1}_2 \bigoplus_{j\in B}({\pm i}) \mathds{1}_2$.

\subsection{Lower bound}

For constructing the lower bound, the matrix $\Gamma=2C-1$ is divided in the following way:
\begin{equation}
\Gamma =\left[
\begin{array}{c@{}|c@{}}
\; \Gamma_{AA} \; & \; \Gamma_{AB} \; \\[2mm]
\hline \\[-4mm]
\Gamma_{AB}^T& \Gamma_{BB}
\end{array}
\right],
\end{equation}
Using the singular value decomposition of $\Gamma_{AB}$, $\Gamma_{AB}=UDV^T$, where $D$ is a diagonal matrix with non-negative elements, whereas $U$ and $V$ are orthogonal matrices, 
one can transform $\Gamma$ by $U\oplus V$ to the following form 
\begin{equation}
\Gamma'=\left[
\begin{array}{c|c}
\; U^T\Gamma_{AA}U \; & \; D \; \\[2mm]
\hline \\[-4mm]
D & V^T\Gamma_{BB}V
\end{array}
\right].
\end{equation}  
Denoting the diagonal elements of $D$, $U^T\Gamma_{AA}U$, and $V^T\Gamma_{BB}V$ by $c_i$, $a_i$, and $b_i$, respectively,
the lower bound is given by  
\begin{equation}
\mathcal{E}_l(\rho_{A \cup B}) =\sum_{j=1}^n \ln h(a_j, b_j, c_j),
\label{lb_gen}
\end{equation}
where
\begin{eqnarray}
h(a,b,c){=} \frac{1}{2}+\frac{1}{2}\max\{1&, \sqrt{(a{+}b)^2 {+}(2c)^2 }{-}(ab{-}c^2) \nonumber \\
&,  |a{-}b|{+}(ab{-}c^2) \}. 
\end{eqnarray}

\subsection{Simplification of the lower bound by sublattice symmetry} 
\label{sec:simple}

Here, we show that the expression in Eq. (\ref{lb_gen}) can be further simplified making use of the sublattice symmetry, which holds in the absence of an on-site potential for an even $L$. We also assume, that the lattice is half filled. Due to this, the elements of matrix $\Gamma=2C-1$ with indices of the same parity are all zero. By replacing rows and columns of $\Gamma$, let us arrange, separately in block $A$ and $B$, the odd (even) indices to the first (second) $l/2$ places. Then $\Gamma_{AB}=2C_{AB}-1$ will have the form
\begin{equation}
\Gamma_{AB}=\left[
\begin{array}{c|c}
\; 0 \; & \; P \; \\[2mm]
\hline \\[-4mm]
Q & 0
\end{array}
\right].
\end{equation}  
We are looking for the singular value decomposition $\Gamma_{AB}=UDV^T$, where the diagonal matrix $D$ contains the non-negative singular values of $\Gamma_{AB}$, while the columns of $U$ and $V$ are eigenvectors of $\Gamma_{AB}\Gamma_{AB}^T$ and $\Gamma_{AB}^T\Gamma_{AB}$, respectively. These matrices are then block diagonal,  
\begin{equation}
\Gamma_{AB}\Gamma_{AB}^T=PP^T\oplus QQ^T \quad 
\Gamma_{AB}^T\Gamma_{AB}=Q^TQ\oplus P^TP.
\label{eq:Gamma_AB}
\end{equation}
and, as a consequence, $U$ and $V$ are block diagonal, as well: 
\begin{equation}
U=U_o\oplus U_e, \quad V=V_o\oplus V_e.
\end{equation}
Transforming $\Gamma$ by $U\oplus V$ will bring the diagonal blocks 
\begin{equation}
\Gamma_{AA}=\left[
\begin{array}{c|c}
\; 0 \; & \; R \; \\[2mm]
\hline \\[-4mm]
R^T & 0
\end{array}
\right], \quad 
\Gamma_{BB}=\left[
\begin{array}{c|c}
\; 0 \; & \; S \; \\[2mm]
\hline \\[-4mm]
S^T & 0
\end{array}
\right],
\end{equation}
to the form
\begin{eqnarray}
\Gamma_{AA}'=&U^T\Gamma_{AA}U=\left[
\begin{array}{c|c}
\; 0 \; & \; U_o^TRU_e \; \\[2mm]
\hline \\[-4mm]
U_e^TR^TU_o & 0
\end{array}
\right], \\ 
\Gamma_{BB}'=&V^T\Gamma_{BB}V=\left[
\begin{array}{c|c}
\; 0 \; & \; V_o^TSV_e \; \\[2mm]
\hline \\[-4mm]
V_e^TS^TV_o & 0
\end{array}
\right].
\end{eqnarray}

The elements $a_i$ and $b_i$ in Eq. (\ref{lb_gen}) are therefore zero.
Using this, the lower bound can be written as   
\be
\mathcal{E}_l=\sum\nolimits_i' 2\ln\left(\frac{c_i+1}{\sqrt{2}}\right),
\label{eq:sing_val}
\ee
where the prime means that the summation goes over the singular values fulfilling $c_i>\sqrt{2}-1$ only.

\section{Numerical results}
\label{results}

We calculated the lower and upper bounds of the entanglement negativity, $\mathcal{E}_l$ and $\mathcal{E}_u$, respectively, which were introduced in Ref. \cite{eisler2018} and are defined in the previous section, for the models described in Sec. \ref{models}.
It is worth mentioning that the above bounds, apart from an additive constant for the upper bound, are sharp for the special case of singlet states. 

The numerical calculations were performed according to two kinds of schemes. 
Either the total system size $L$ was kept fixed (and large) and the size $\ell$ of adjacent subsystems was varied, or $L$ was varied keeping the ratio $\ell/L$ constant. For a given $L$ and $\ell$, $\mathcal{E}_l$ and $\mathcal{E}_u$ were calculated for all possible positions of the subsystems ($L$ in number) and an averaging was performed in the case of non-random models. 
For the random model, $32$ different positions of the subsystems was considered in each random sample, while the number of samples was $10^6$ for smaller systems, gradually decreasing down to $5000$ for the largest system with $L=2048$.        

In order to make corrections to an expected large-$L$ dependence
\be
\mathcal{E}_{l,u}(L)=\kappa_{l,u}\ln L + {\rm const}
\label{EL}
\ee
more visible in the second scheme, we also calculated effective, size-dependent 
prefactors from consecutive data points at $L$ and $L'>L$ as 
\be 
\kappa_{l,u}(L)=\frac{\mathcal{E}_{l,u}(L')-\mathcal{E}_{l,u}(L)}{\ln(L'/L)}.
\label{kappa_eff}
\ee

For the Harper model, we also considered the average entanglement entropy of a subsystem of size $\ell$, the scaling of which is so far unexplored, and which can be numerically calculated from the eigenvalues of the correlation matrix of fermions restricted to the subsystem \cite{vidal,peschel-2003}.  

\subsection{Homogeneous system}
\label{subsec:hom}

\begin{figure}
\begin{center}
\includegraphics[width=75mm, angle=0]{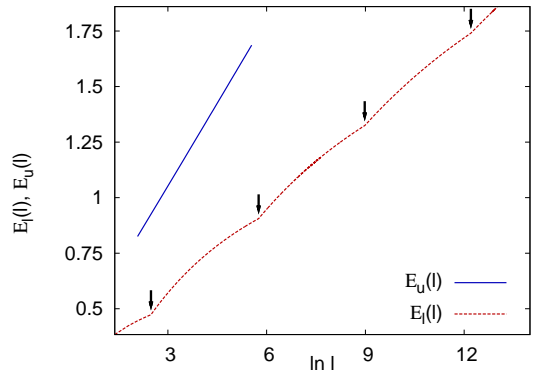}
\end{center} 
\caption{
\label{hom-XX}
Upper and lower bounds for the entanglement negativity in the homogeneous fermion chain. The subsystem sizes are up to $\ell=1024$ in the case of the upper bound, and  $\ell=500000$ in the case of the lower bound. 
The lower bound shows log-periodic oscillations; the boarders of the periods are indicated by arrows.
	}
\end{figure}

First, we investigated the behavior of the lower and upper bounds in the homogeneous chain. Here, we assume that the system is infinite ($L \to \infty$), which allows us to use the exact expressions of the correlations $C_{i,j}=\langle c^{\dagger}_i c_j \rangle$ \cite{igloi2020}, while keep the subsystem size $\ell$ finite. The numerical results are shown in Fig. \ref{hom-XX}.
The upper bound follows the scaling 
\begin{equation}
\mathcal{E}_u(\ell) \sim \kappa_u \ln\ell + \mathrm{const}
\end{equation} 
with  $\kappa_u=0.25 \pm 0.0005$, which is compatible with the behavior of the entanglement negativity known from CFT, see Eq. (\ref{eq:lnl}). 

The behavior of the lower limit is more complicated: the overall logarithmic trend is decorated with log-periodic oscillations of small amplitude, which is unusual in a homogeneous system. Here, the oscillations are the consequences of the definition of the lower limit [see Eq. (\ref{eq:sing_val})], in which only the singular values greater than $\sqrt{2}-1$ contribute to the sum. Putting the singular values in decreasing order, one can observe that the $n$th singular value ($n=1,2,\dots$) slowly increases with increasing $\ell$. When a singular value crosses the threshold $\sqrt{2}-1$, so that the number of terms in the sum increases by one, a new period of the oscillations starts.
  
To investigate the overall logarithmic trend, one has to fit to identical parts of the periods, like the breaking points indicated by arrows in Fig. \ref{hom-XX}. To do this, at least a few period long data set is needed. The period of these oscillations is about $3.2$ on a logarithmic scale, which means that one has to handle quite large subsystems of size $10^5 \dots 10^6$. 
Fortunately, in this size range, only the largest $1 \dots 4$ singular values are needed, which can be obtained from the largest eigenvalues of the matrices
 $QQ^T$ and $P^TP$. Since we have a closed formula for the matrix elements of $Q$ and $P$, we can use the power method and multiplicate with $QQ^T$ ($P^TP$) without storing the matrix elements in the memory. 
The ratios of the consecutive eigenvalues (in decreasing order) of the matrix $Q Q^T$ ($P^T P$) are between $0.25$ and $0.1$, so the power method converges rapidly.

Using the data for the lower bounds at the four breaking points shown in Fig. \ref{hom-XX}, we calculated effective prefactors from neighboring data points, which are in order, $0.1321$, $0.1298$, and $0.1288$. 
Assuming corrections of the form $1/\ln\ell$, an extrapolation to $\ell\to\infty$ results in $\kappa_l=0.1256$; if the data point obtained from the smallest system sizes is excluded, one obtains $0.1246$, which gives an estimate on the error of the fit. 
Thus, in the homogeneous chain, the prefactor of the lower bound, $\kappa_l=0.125(1)$, is close to the half of the prefactor of entanglement negativity $1/4$.  

\subsection{Off-diagonal inhomogeneity}

\begin{figure}
\begin{center}
\includegraphics[width=75mm, angle=0]{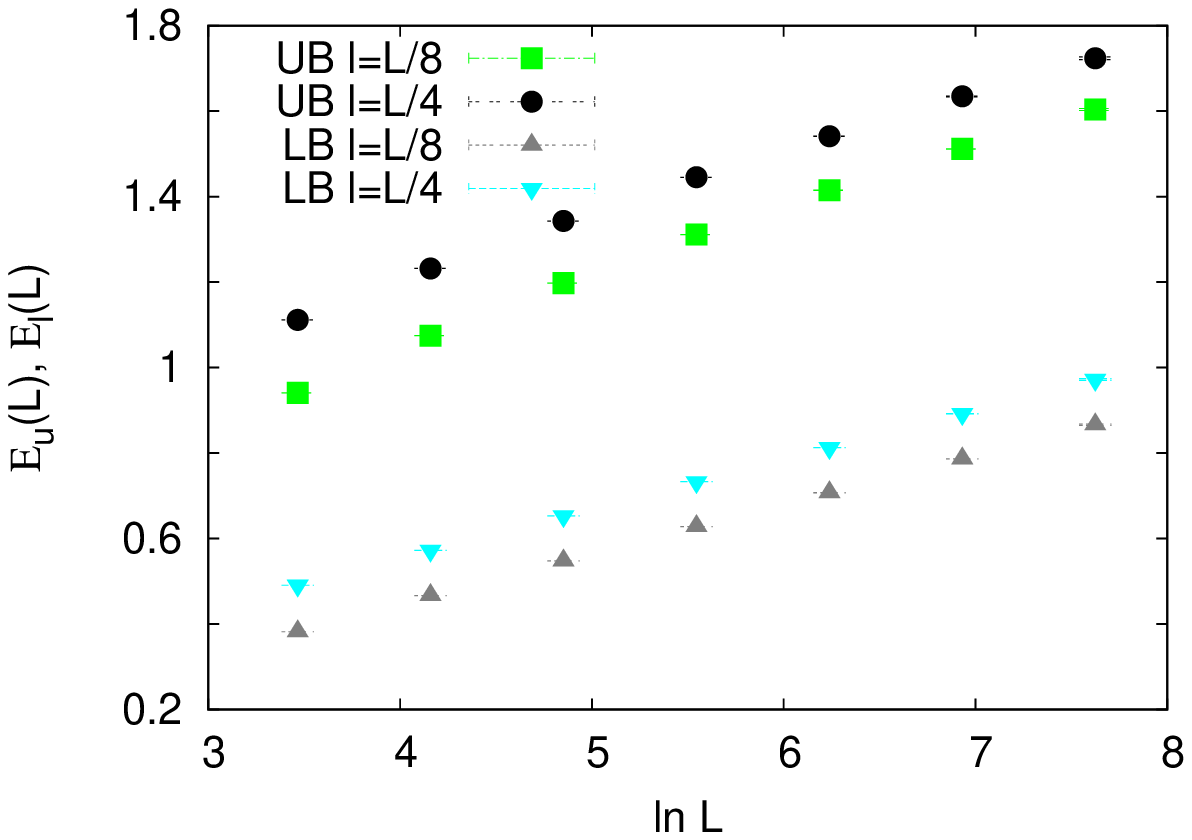}
\includegraphics[width=75mm, angle=0]{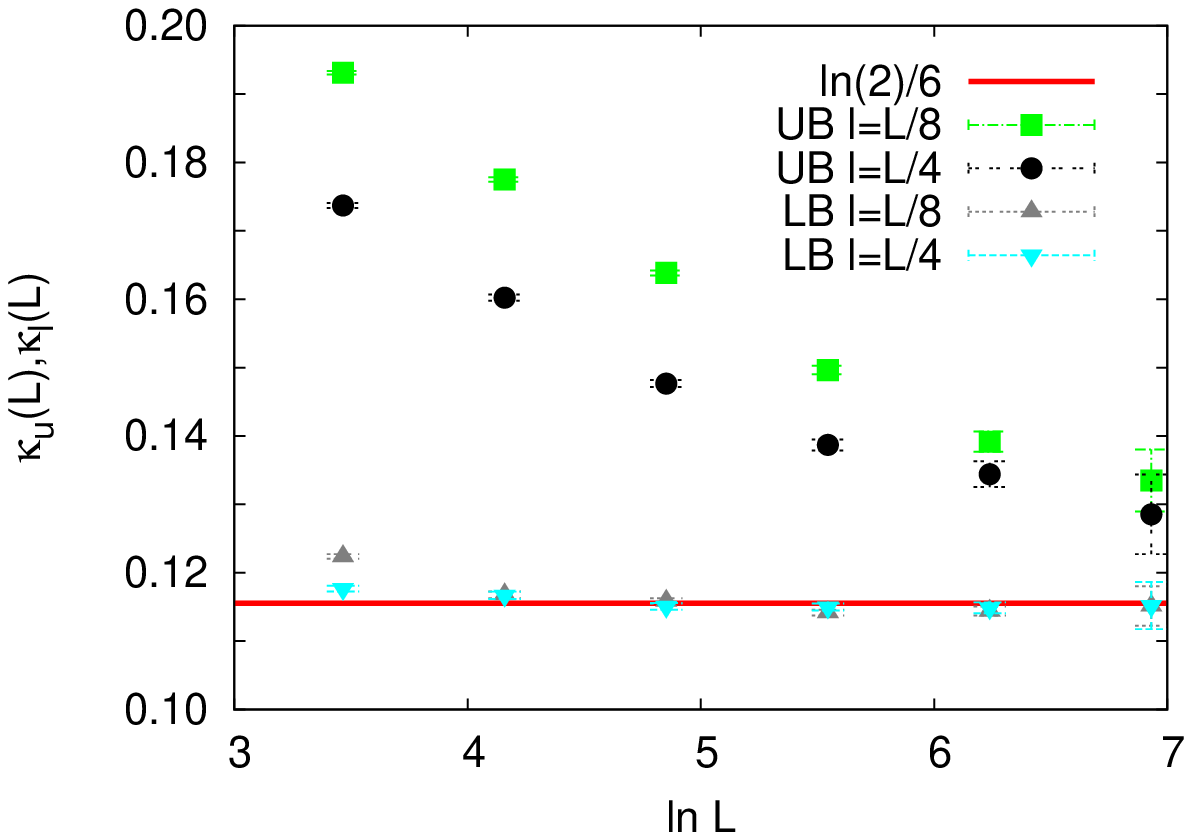}
\end{center} 
 \caption{
\label{random_fig}
Left. Lower (LB) and upper (UB) bound of the entanglement negativity in the random model, as a function of the system size $L$, for fixed ratios $\ell/L=1/8$ and $\ell/L=1/4$. 
Right. The corresponding effective prefactors defined in Eq. (\ref{kappa_eff}). The horizontal line indicates the asymptotic value $\ln(2)/6=0.1155\dots$ predicted by the SDRG method. 
}
\end{figure}

The dependence of the lower and upper bounds on $L$ for fixed ratios $\ell/L$, as well as the corresponding effective prefactors for the random model are shown in Fig. \ref{random_fig}.  
As can be seen, both the lower and upper bounds scale logarithmically with $L$ for large $L$. The effective prefactor of the upper bound displays a slow crossover from the clean system's value $1/4$ toward $\ln(2)/6$ predicted by the SDRG method with increasing $L$, and, at the system sizes available by the numerical method, it is still considerably far from it. We note that, under the same circumstances, the prefactor of the entanglement entropy has similar deviations from the asymptotic value (not shown).
  In the case of the lower bound, the prefactor in the homogeneous system ($0.13$) is closer to the asymptotic limit $\ln(2)/6$, and, accordingly, it shows a more rapid crossover.   
  
\begin{figure}
\begin{center}
\includegraphics[width=8cm, angle=0]{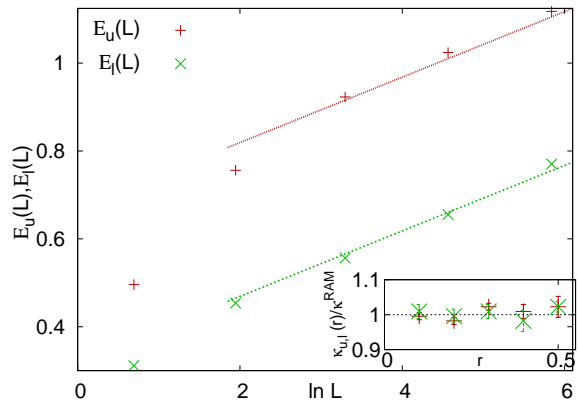}
\end{center} 
 \caption{
\label{ram_fig}
Left. Lower and upper bound of the entanglement negativity for RAM, as a function of the system size $L$. 
Inset. The ratios of the prefactors $\kappa_{u,l}(r)$ obtained for different values of $r$ and $\kappa^{\rm RAM}=c_{\rm eff}^{\rm RAM}/6=0.07432\dots$ predicted by the SDRG method. 
}
\end{figure}

In the case of the relevant aperiodic modulation, the lower and upper bounds, as well as the effective prefactors as a function of $L$ are exemplified in Fig. \ref{ram_fig} for $r=0.5$. 
The system sizes were $L=17,61,217,773,2753$, which are the lengths of words obtained by the repeated application of the inflation rule starting with letter $a$, while the subsystem size was $\ell=[L/8]$, where $[\cdot]$ stands for the integer part. By this choice of $L$, log-periodic oscillations in the data can be avoided \cite{jz}.  
As can be seen, the bounds follow a logarithmic scaling and their asymptotic prefactors shown in the inset for different disorder strengths $r$ are in agreement with the expectation $\kappa_{l,u}=\kappa=c_{\rm eff}/6$ valid for singlet states. 

\begin{figure}
\begin{center}
\includegraphics[width=9cm, angle=0]{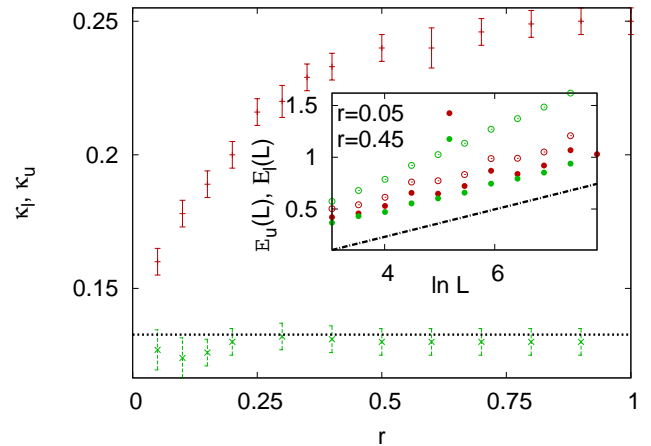}
\end{center}
 \caption{
\label{Fib_fig} Prefactors of the lower and upper bounds as a function of the modulation strength $r$ for the Fibonacci modulation. 
The horizontal line indicates the singlet-state limit, $\lim_{r\to 0}c_{\rm eff}^{\rm FM}(r)/6=0.1327\cdots$. In the inset, the dependence of the upper (open symbols) and lower (full symbols) bound on $L$ for two different values of the modulation strength are shown. The straight line has a slope characteristic for the singlet-state limit.}
\end{figure}

For the Fibonacci modulation, which is a marginal one, we calculated the size dependence of lower and upper bounds for different modulation strengths $r$. 
Here, the system sizes and the subsystem sizes were chosen to be  $L=F_n$ and $\ell=F_{n-4}$, respectively, where $F_n$ denotes the $n$th term of the Fibonacci sequence. 
As can be seen in Fig. \ref{Fib_fig}, the bounds plotted against $\ln L$ show log-periodic oscillations with a period of three data points. This is in accordance with that the self-similarity of the aperiodic singlet state is achieved by applying the third power of the inflation transformation \cite{jz}. 
Therefore, in estimating the prefactors, we used every third data points.   
For both bounds, we find a logarithmic dependence on $L$ as given in Eq. (\ref{EL}). The prefactor of the upper bound shows a variation with $r$ qualitatively similar to that of the prefactor of the entanglement entropy \cite{jz}: 
For moderate strengths of the modulation, it has a slight deviation from the clean system's value, then it tends rapidly to the singlet-state limit.  
In the case of the lower bound, the prefactor in the clean system ($0.125(1)$) and in the singlet-state limit ($0.1327\dots$) hardly differ, therefore the variation with $r$ is very weak.  

\subsection{The Harper model}

\begin{figure}
	\begin{center}
		\includegraphics[width=75mm, angle=0]{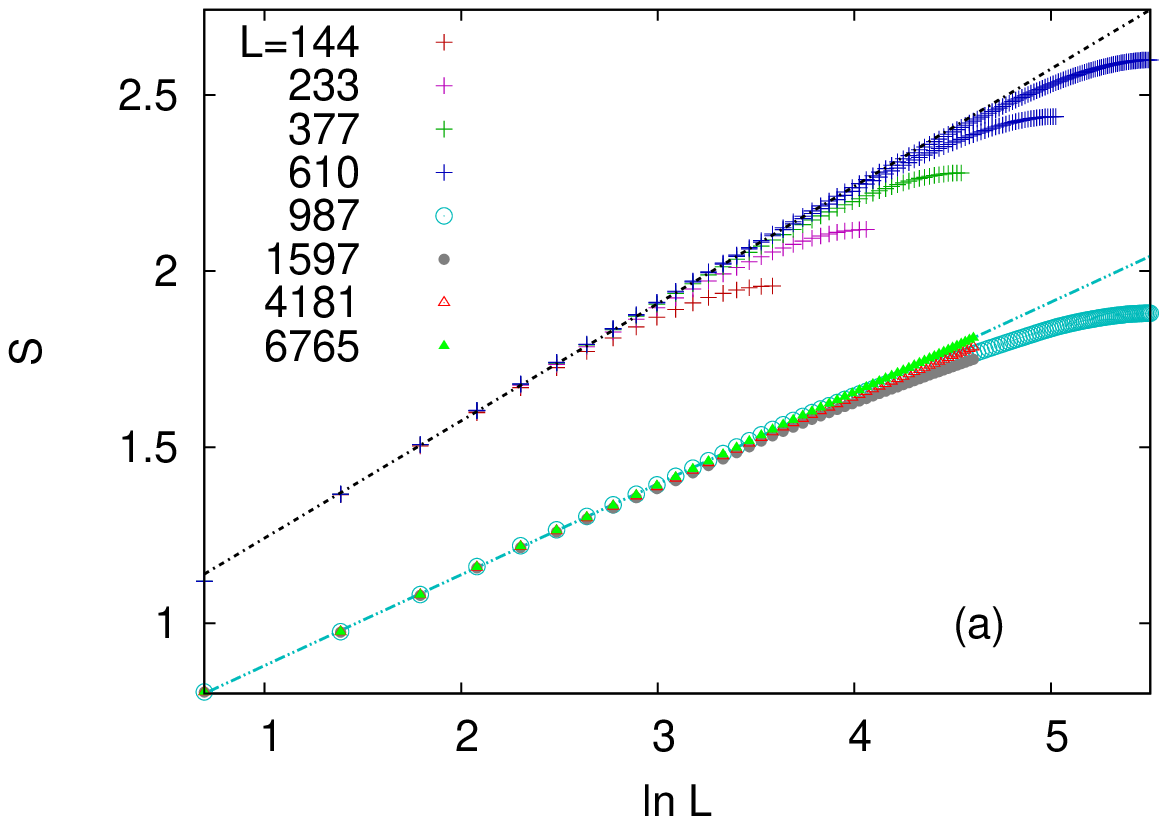}
		\includegraphics[width=75mm, angle=0]{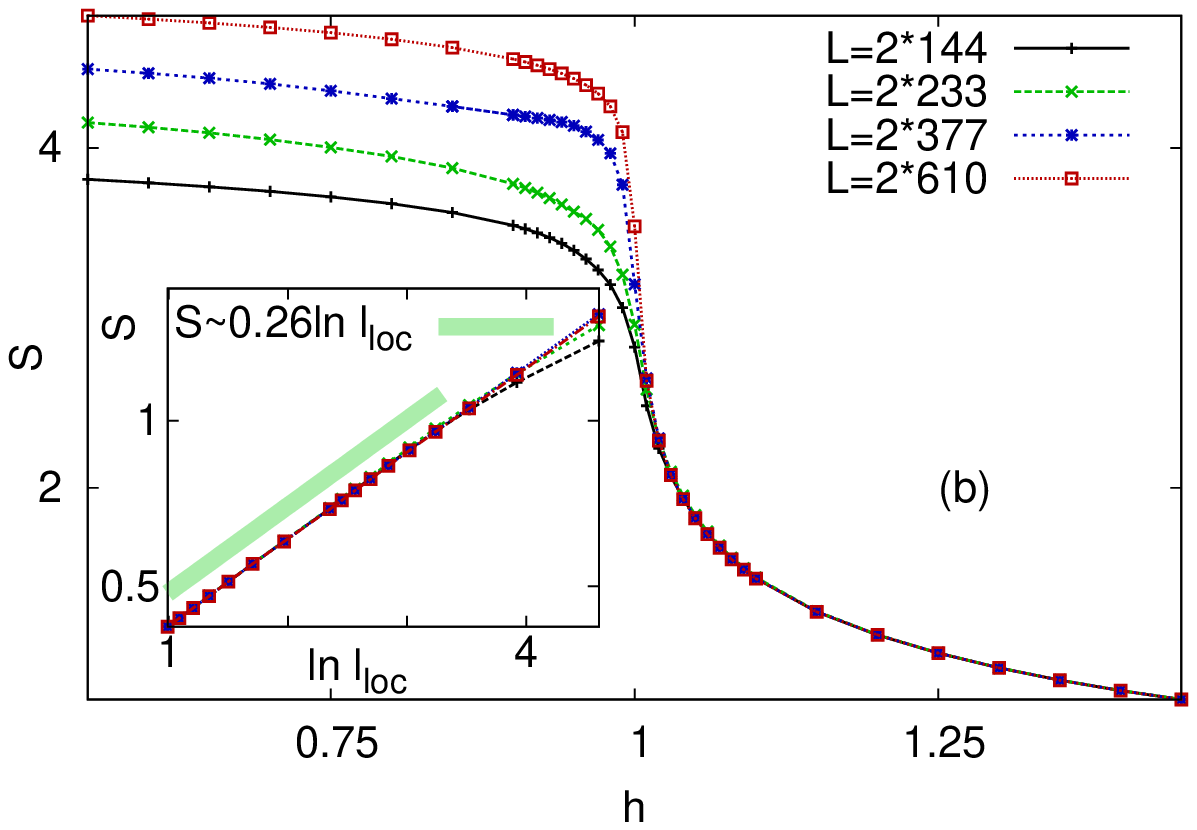}
	\end{center}
	
	\caption{\label{fig_harper_entanglement_entropy}  Entanglement entropy of the Harper model. a) Entanglement entropy in the extended phase $h=0.5$ (upper data) and in the critical point $h=1.0$ (lower data) as a function of $\ln L$. The straight lines have slopes $0.33$ and $0.26$.
b) Entanglement entropy plotted against $h$ for various system sizes. The inset shows the entanglement entropy in the localised phase as a function of the logarithm of the localization length $l_{\rm loc}=1/\ln h$. The green line has a slope $0.26$.} 
\end{figure}

In the Harper model, we investigated first the scaling of the entanglement entropy in the ground state.  
In these calculations, system sizes were chosen to be twice of Fibonacci numbers $L=2 F_n$, and the size of the subsystem was $F_{n-4}$. 

The results are shown in Fig. \ref{fig_harper_entanglement_entropy}. 
We find that, in the extended phase, the entanglement entropy scales logarithmically as $S = 0.33\ln L + \textnormal{const}$, and the measured prefactor ($0.33$) is compatible with that of the homogeneous system $1/3$ up to the precision of the estimation.  
The reason of this agreement is the extended nature of the eigenstates, which is similar to the eigenstates of a homogeneous chain.

In the critical point, the entanglement entropy is still found to scale logarithmically as
\begin{equation}
 S(L) = \frac{c_{\rm eff}}{3}\ln L + \textnormal{const},
\label{ceff}
\end{equation}
however, the effective central charge $c_{\rm eff}=0.78$ differs significantly from the central charge of the homogeneous system.

In the localized phase, the entanglement entropy saturates to finite values in the limit $L\to \infty$. As it is demonstrated in Fig. \ref{fig_harper_entanglement_entropy}, in large systems $L\gg l_{\rm loc}$, in which the length scale is set by the localization length $l_{\rm loc}$ rather than the system size, the entanglement entropy follows the law
\begin{equation}
 S(l_{\rm loc}) = \frac{c_{\rm eff}}{3}\ln l_{\rm loc} +  \textnormal{const.}
\end{equation}
with the same prefactor as found in Eq. (\ref{ceff}).  
Approaching the critical point, this law is, however, deteriorates when the diverging localisation length becomes comparable with the system size.

\begin{figure}[t]
\begin{center}
\includegraphics[width=75mm, angle=0]{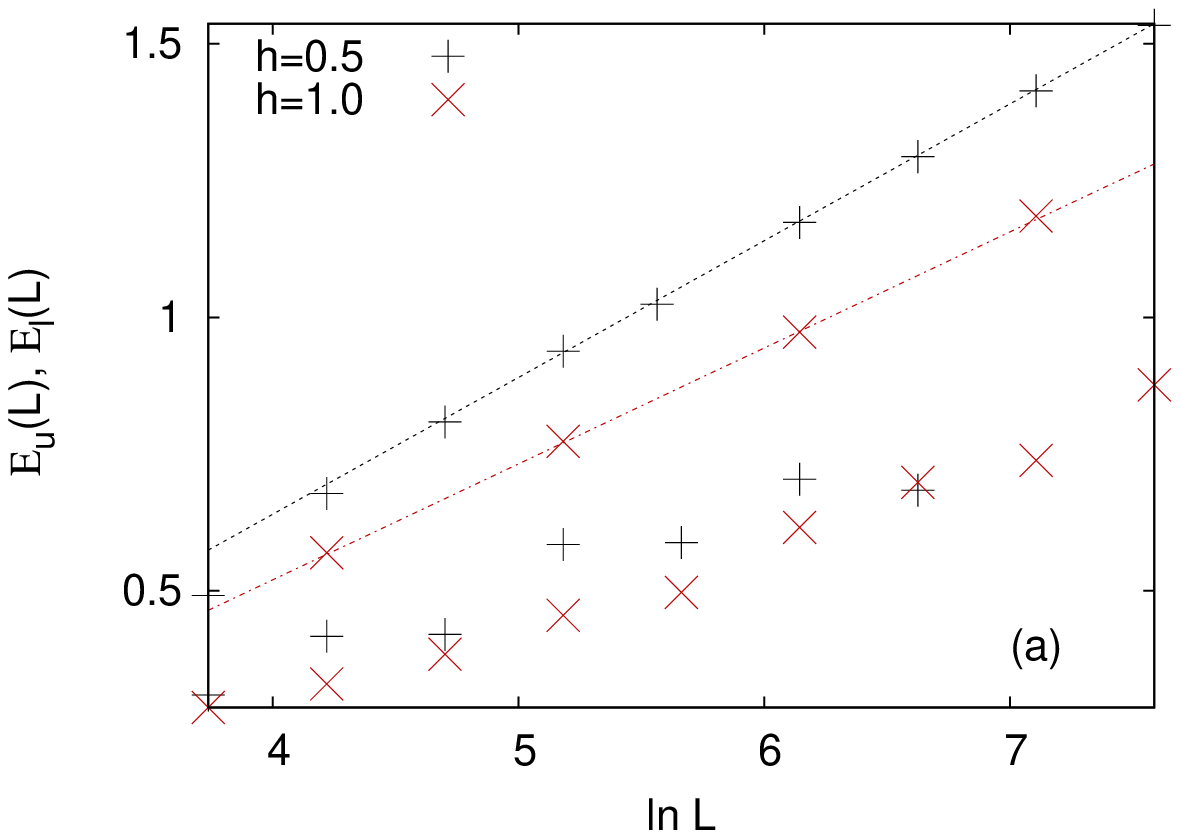}
\includegraphics[width=75mm, angle=0]{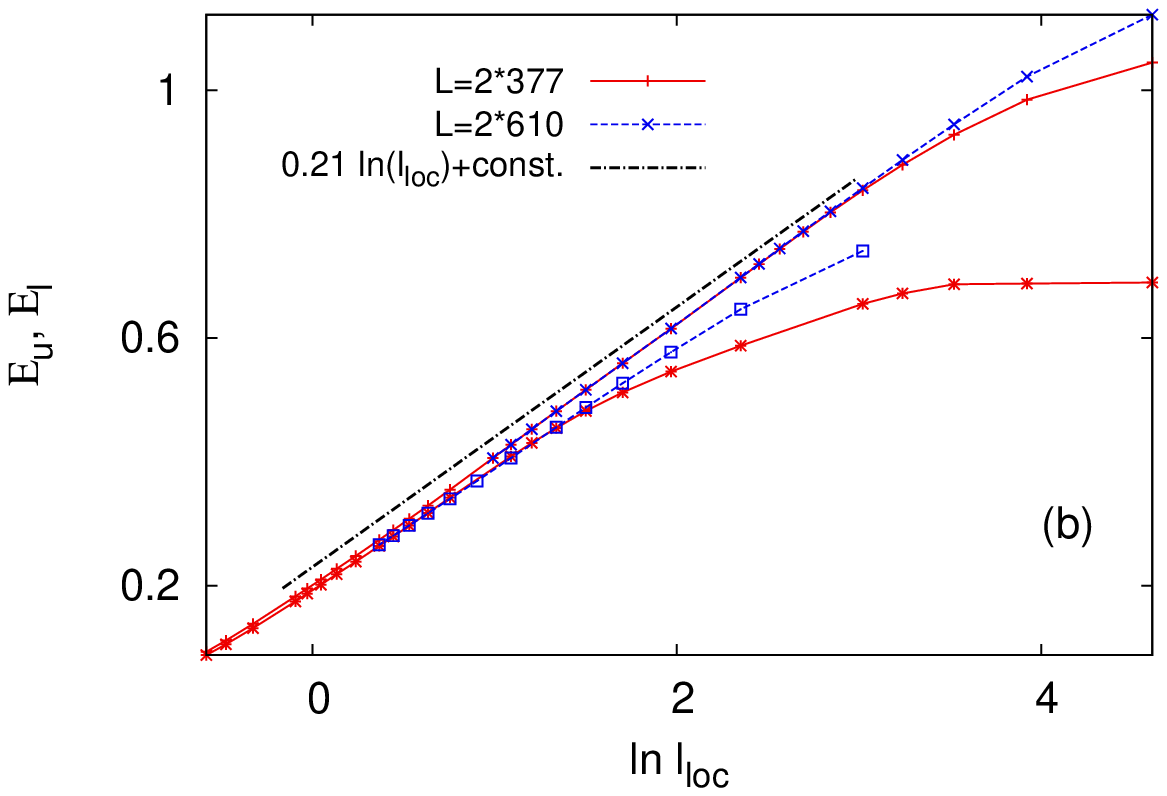}
\end{center}
 \caption{Upper and lower bounds of the entanglement negativity of the Harper model. a) Dependence on the system size in the delocalized phase ($h=0.5$) and in the critical point ($h=1$). The slopes of fitted lines from top to bottom are $0.25$, $0.212$, $0.16$, and $0.16$
 \newline
 b) Dependence on the localization length in the localized phase. The black line with slope $0.21$ is drawn to guide the eye.  
   \label{FIG5}}
\end{figure}

Numerical results on the entanglement negativity are shown in Fig. \ref{FIG5}.
In the delocalized phase and in the critical point, the upper bound increases logarithmically with the system size. In the delocalized phase, just like for the entranglement entropy, the prefactor is found to agree with the that of the conformally invariant homogeneous system. 
In the critical point, the prefactor of the upper bound is reduced to $\kappa_u=0.212$.  As can be seen in Fig. \ref{FIG5}, the size dependence of the lower bound shows oscillations which may originate both in the discreteness in its definition (see section \ref{subsec:hom}) and in the quasiperiodic modulation. For this reason, a completely reliable estimate of the prefactor cannot be obtained.
 
As can seen in Fig. \ref{FIG5}b, the upper and lower bounds become nearly equal in the localized phase.
Their ratio approaches to one, if $h\to \infty$, and the ratio is $1.05$  already for $h=1.5$. This makes possible to investigate the behavior of the 
negativity in the localized phase. 
Although the difference of the upper and lower bounds becomes larger as one approaches the transition point, there is an intermediate regime, where the transition point is not too far, but the difference of the bounds is still moderate. 
Similarly to the behavior of the entanglement entropy in the localized phase, the upper bound of entanglement negativity increases according to
\begin{equation}
\mathcal{E}_{u}(l_{\rm loc}) = \kappa_{u}\ln l_{loc} + const.
\end{equation}
in the vicinity of the transition point, where $\kappa_{u}=0.212$ is the prefactor characteristic of the critical point.

\section{Conclusion and outlook}
\label{concl}

In this paper, we investigated numerically the scaling of the entanglement negativity in fermionic chains with different aperiodic and random modulations.
In general, we found a logarithmic increase of both bounds with the subsystem size, provided the model is critical. 
Some of the off-diagonal modulations, like the random, the relevant aperiodic, and the extremal Fibonacci modulation asymptotically induce a state composed of singlets. In these cases, we have confirmed that the prefactors of the lower and upper bounds coincide and agree with the half of the prefactor of the entanglement entropy, $\kappa=\frac{c_{\rm eff}}{6}$.

In the case of the Fibonacci modulation of finite strength $r$, the ground state is no longer a singlet state, and the prefactor of the entanglement entropy is known to vary continuously and monotonically  with $r$. We have found a similar behavior of the prefactors of both bounds, although for the lower bound the difference between the two extremal values is very small. 
The prefactor of the upper bound, which correctly gives the prefactor of the entanglement negativity in the extremal cases $r\to 0$ and $r=1$, is presumably holds to be correct in the intermediate regime $0<r<1$, as well.     
However, $\kappa(r)$ is not expected to be simply related to $c_{\rm eff}(r)$, as their ratio is different in the two extremal cases ($1/2$ and $3/4$).  
          

The Harper model, which contains a quasi-periodic modulation in the diagonal term, has different ground-state phases depending on the modulation strength. 
In the extended phase, the entanglement entropy and the upper bound of the negativity are found to  scale identically with the conformally invariant homogeneous chain. This is in accordance with the expectations since, in this phase, the Harper model has extended quasi-free eigenstates a continuous spectrum \cite{sinai1987}.
In the critical point separating the extended phase from the delocalized one, we find, however, that the prefactors of both the entanglement entropy and the upper bound of the negativity are reduced compared to the values of the homogeneous system. It is remarkable that the prefactors are reduced by roughly the same factor ($f\approx 0.78$ for the entanglement entropy and $f\approx 0.85$ for the negativity upper bound). This may suggest that the scaling can be described by the prefactors of the homogeneous system, but using an effective length $L'\sim L^{f}$.   
Indeed, the one-particle states of the model in the critical point are known to be fractals, with an effective number of sites participating in them scaling as $N\sim L ^{D_2}$. The fractal dimension $D_2$ varies from state to state and  its maximal value\cite{evangeleou2000} $D_2\approx 0.82$  is close to $f$.  
The clarification of a possible relationship between the reduction factor and the fractal dimensions of eigenstates is an open question which is deferred to future research.

As a possible future direction, the present investigations could be extended to  various generalized Harper models. For instance, one could add paring terms ($c_l c_{l+1}$) to the Hamiltonian and investigate the so called quasi-periodic Ising  model \cite{chandran2017}  with the free-fermion methods applied here. We also mention here recent generalizations of the  Harper model with additional long-range couplings \cite{monthus2019, deng2019}.

\begin{acknowledgments}
This work was supported by the by the Deutsche Forschungsgemeinschaft through the Cluster of Excellence on Complexity and Topology in Quantum Matter ct.qmat (EXC 2147) and the National Research, Development and Innovation Office NKFIH under grants No. K128989, K124152, K124176, KH129601 and through the Hungarian Quantum Technology National Excellence Program (Project No. 2017-1.2.1-NKP-2017-00001). ZZ was also partially funded by the Janos Bolyai and the Bolyai+ Scholarships. 
\end{acknowledgments}

\end{document}